\documentclass[aps,prb,twocolumn,superscriptaddress]{revtex4}%
\usepackage{epsfig,dsfont,amssymb,amsmath,amsthm,amsfonts,amsbsy,mathrsfs}
\usepackage{graphicx}
\usepackage{color}
\usepackage{epstopdf}
\usepackage{float}
\usepackage{amsmath}
\usepackage{amssymb}%
\begin{document}
\title{Quantum phase transition in a multiconnected Jaynes-Cummings lattice}
\author{Jian Xue}
\affiliation{Institute of Physics, Chinese Academy of Sciences, Beijing 100190, China}

\author{Kangjun Seo}
\affiliation{School of Nature Sciences, University of California, Merced, California 95343, USA}

\author{Lin Tian}
\email{ltian@ucmerced.edu}
\affiliation{School of Nature Sciences, University of California, Merced, California 95343, USA}

\author{Tao Xiang}
\email{txiang@iphy.ac.cn}
\affiliation{Institute of Physics, Chinese Academy of Sciences, Beijing 100190, China}
\affiliation{Collaborative Innovation Center of Quantum
Matter, Beijing 100190, China}

\begin{abstract}
  The rapid progress in quantum technology enables the implementation of artificial many-body systems with correlated photons and polaritons. A multiconnected Jaynes-Cummings (MCJC) lattice can be constructed by connecting qubits and cavities alternatively. Such kind of models can be realized with superconducting qubits coupled to superconducting microwave resonators or with quantum dots coupled to optical nanocavities. We study physical properties of the one-dimensional MCJC lattice using the density-matrix renormalization group method. This model has an intrinsic symmetry between the left and right qubit-cavity couplings. The competition between these couplings may drive the ground state either to a Mott-insulating or to a superfluid phase at an integer filling. We calculate the single-particle and density-density correlation functions, the correlation lengths in the Mott-insulating phase and the Luttinger parameters in the superfluid phase, and determine accurately the critical points that separate these two phases.
\end{abstract}
\maketitle

\section{Introduction}
The past few decades have witnessed enormous progress in the development of quantum devices in various physical systems, such as superconducting devices, trapped ions, and semiconductor photonic devices, with significant improvement in their controllability and coherent property.~\cite{squbit_review, ion_review, semiconductor_review, photonics_review} Besides the goal of building scalable fault-tolerant quantum computers,~\cite{qc_review} these devices have been exploited to
emulate numerous many-body phenomena that are difficult to solve with conventional techniques~\cite{Feynman, LloydQS, HaukeReview, atom_review1, atom_review2}
in condensed-matter physics, high-energy physics, and nonequilibrium systems.~\cite{AbramsPRL1997, LAWu2002, qchem1, squbit_exp4, eph1, eph2, Mei:2013, Stojanovic:2014, squbit_exp1, squbit_exp2, Tian:2010, SolanoPRL2014, squbit_exp5, Kapit:2013, Marcos:2013, PeropadrePRB2013}

The construction of artificial many-body systems lead to the study of strongly-correlated photons and polaritons.~\cite{Hartmann:2008, Houck:2012} In the coupled cavity array (CCA) models,\cite{Hartmann:2006, Greentree:2006, Angelakis:2007} photons can hop between adjacent cavities. The cavity modes are also coupled to nonlinear medium, such as qubits and defects, which adds nonlinearity to the spectrum of the polariton excitations. The nonlinearity can be viewed as an onsite Hubbard interaction, in comparison to the Bose-Hubbard (BH) model.~\cite{Fisher:1989, Batrouni:1990, KuhnerPRB1998, KuhnerPRB2000} The competition between the hopping and the nonlinearity results in quantum phase transitions between the Mott-insulating (MI) phase with localized polariton excitations and the superfluid (SF) phase with long-range spatial correlation. In the past few years, the CCA has been studied extensively in theory and in experiments.~\cite{RossiniPRL2007, Na:2008, MakinPRA2008, SchmidtPRL2009, Koch:2009, PippanPRA2009, GreentreePRL2012, DSouzaPRA2013, Srinivasan:2011} Photon blockade has been demonstrated in a recent experiment.~\cite{Hoffman:2011} Dynamical quantum phase transition with driven and dissipative cavities has been investigated.~\cite{KeelingPRL2012, Houck2016}  In recent works,~\cite{Seo2015:1, Seo2015:2, QiuPRA2014, GarciaRipollPRL2014} a multiconnected Jaynes-Cummings (MCJC) lattice was introduced, where qubits and cavities are connected alternatively. Both CCA and MCJC can be realized with the microwave modes in superconducting resonators coupled to superconducting qubits or with optical modes in nanocavities coupled to quantum dots or defects.~\cite{squbit_review, semiconductor_review} A specific realization of the MCJC lattice is to connect Xmon qubits to superconducting resonators, enabled by the rich connectivity of superconducting circuits.~\cite{Barends:2013, Chen:2014} Different from the CCA, no direct coupling exists between cavities in the MCJC. Quantum phase transition in a MCJC lattice has been studied with exact diagonalization.~\cite{Seo2015:1, Seo2015:2} It was shown that at integer filling, transitions between the MI and SF phases occur due to the competition between  the qubit-cavity couplings. These systems provide a promising platform to study correlations in interacting photons and polaritons.

In this paper, we study the quantum critical behavior of the one-dimensional (1D) MCJC model using the density-matrix renormalization group (DMRG).~\cite{White1992, White1993} This method has previously been used to study the BH and CCA models.~\cite{KuhnerPRB1998, KuhnerPRB2000, RossiniPRL2007, DSouzaPRA2013} We calculate the polariton ground states at both integer and half fillings. The phase boundaries separating the MI and SF phases are obtained by calculating the chemical potentials.~\cite{Sachdev} The single-particle density matrix is utilized to obtain the correlation lengths and the Luttinger parameters of the qubits and the cavities. Using the Luttinger parameters extrapolated to the thermodynamic limit, the quantum critical points are determined accurately and compared with the previous results.~\cite{Seo2015:1, Seo2015:2} We also calculate the structure factors, and find no evidence for a crystalline or charge-density-wave (CDW) phase at half filling. Our result may shed light on the study of strongly-correlated photons and polaritons, especially stationary and out-of-equilibrium effects in the MCJC and related models.

The paper is organized as follows. In Sec.~II, we introduce the MCJC model and analyze its phase diagrams at integer and half fillings. In Sec.~III, we describe the DMRG method and discuss the results of entanglement entropy and local density. In Sec.~IV, the DMRG results on the phase boundary, single-particle density matrix, correlation length, Luttinger parameter, and density-density correlation function are discussed. Conclusions are given in Sec.~V.

\section{Multiconnected Jaynes-Cummings lattice}

\begin{figure}
\centering
\includegraphics[width=8cm, clip]{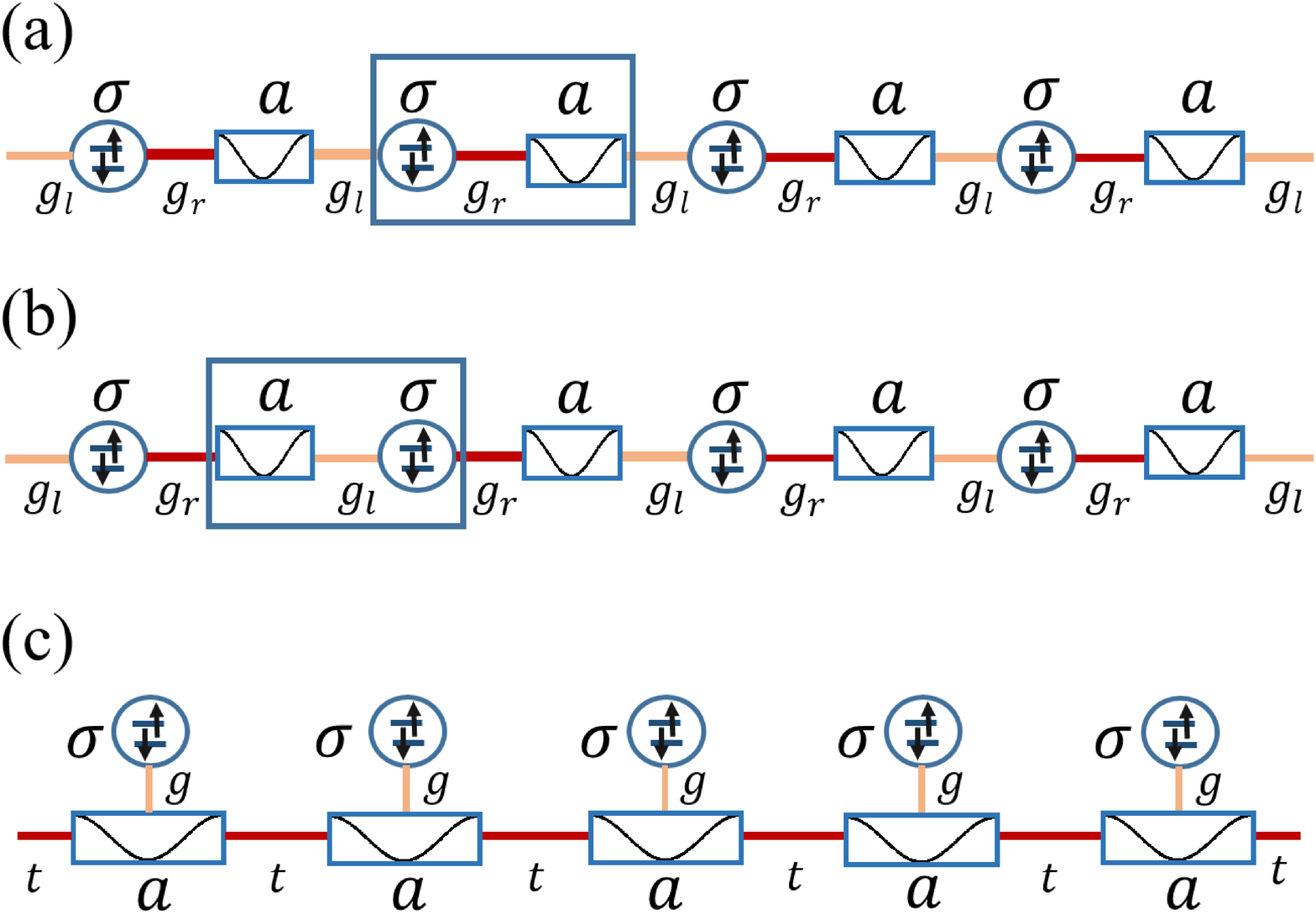}
  \caption{Schematic of a 1D MCJC lattice (a, b) and a CCA (c). The circles represent the qubits and the blocks represent the cavities. A unit cell of the MCJC consists of a qubit and a cavity adjacent to it coupled (a) by $g_r$ or (b) by $g_l$.}
\label{fig:mcjc}
\end{figure}

\subsection{The model}

A MCJC lattice is composed of alternatively connected qubits and cavities.~\cite{Seo2015:1, Seo2015:2} A schematic of a one-dimensional (1D) MCJC lattice is shown in Fig.~\ref{fig:mcjc}, where each qubit couples to two adjacent cavities. The Hamiltonian of this model reads ($\hbar=1$)
\begin{eqnarray}
H_{t} &=& H_0+H_{l} + H_{r}, \label{eq:Ht}  \\
H_0 &=& \sum_{i} \left(\frac{\omega_z}{2}{\sigma}^z_{2i-1}+\omega_c {a}^\dagger_{2i}{a}_{2i}\right), \label{eq:H0} \\
H_{l} &=& \sum_{i} g_l \left({\sigma}^{+}_{2i-1}{a}_{2i-2}+{a}^\dagger_{2i-2}{\sigma}^{-}_{2i-1}\right), \label{eq:Hl}\\
H_{r} &=& \sum_{i} g_r\left({\sigma}^{+}_{2i-1}{a}_{2i}+{a}^\dagger_{2i}{\sigma}^{-}_{2i-1}\right), \label{eq:Hr}
\end{eqnarray}
where $\omega_z$ is the energy splitting of the qubits, $\omega_c$ is the frequency of the cavity modes, ${\sigma}_{i}^{z,+,-}$ are the Pauli matrices, ${a}_i$ is the annihilation operator of cavity modes, $g_l$ ($g_r$) is the coupling constant between a qubit and a cavity next to it from the left (right) hand side.
In this paper, periodic boundary conditions are assumed.

Similar to the CCA model [Fig.~\ref{fig:mcjc}(c)], a  unit cell contains one qubit and one cavity. But unlike in the CCA model, the cavities are not directly coupled to each other. The unique geometry of this model renders a symmetry between $g_l$ and $g_r$, and the energy spectrum is unchanged under the exchange of these two coupling constants. This leads to the symmetry in the phase diagram discussed below.

The energy spectrum of a single unit cell in the MCJC lattice can be exactly solved. If we take a qubit with its right cavity as a unit cell [Fig.~\ref{fig:mcjc}(a)], the Hamiltonian of the $i$th unit cell is given by
\begin{equation}
H_i=\omega_c {a}_{2i}^\dagger {a}_{2i}+\frac{\omega_z}{2}{\sigma}^z_{2i-1}+g_r\left({a}_{2i}^\dagger {\sigma}_{2i-1}^-+{\sigma}_{2i-1}^+ {a}_{2i}\right),  \label{eq:H0i'}
\end{equation}
which is nothing but the standard Jaynes-Cummings (JC) model between a qubit and a cavity.~\cite{JCmodel} It has been extensively studied in cavity and circuit quantum electrodynamic (QED) systems.~\cite{cavityQED, circuitQED1, circuitQED2} In the ground state, there is no photon excitation and the qubit is in the down-spin state. If we use $|n,\sigma\rangle$ to denote the basis states, then the ground state is given by $|0,\downarrow\rangle$,
where $n$ is the number of photons in the cavity and $\sigma=(\uparrow, \downarrow )$ corresponds to the two spin states of the qubit, with ground-state energy $E_{0}=-\omega_z/2$.
The higher excitation states appear in pairs and can be regarded as a doublet, resulting from the coupling between the basis states $|n-1,\uparrow\rangle$ and $|n,\downarrow\rangle$. In this doublet subspace, the Hamiltonian can be expressed as a 2$\times 2$ matrix
\begin{equation}
\left[H_i\right]_{n}=
\left(\begin{array}{cc}
(n-1)\omega_c+\frac{\omega_z}{2}    &\sqrt{n}g_r\\
\sqrt{n}g_r & n\omega_c-\frac{\omega_z}{2}
\end{array}\right) . \label{eq:H0i'n}
\end{equation}
There is no coupling between different doublets. Diagonalizing this matrix, we obtain the eigenstates
\begin{equation}
|n,\pm\rangle=\gamma_{n\pm}|n,\downarrow\rangle + \rho_{n\pm}|n-1,\uparrow\rangle ,
\label{eigenstates}
\end{equation}
where $\gamma_{n +}= - \rho_{n - }= \sin \theta_n$, $\gamma_{n -}= \rho_{n +} = \cos \theta_n$, and

\begin{equation}
\theta_n=\arctan\left(\frac{\sqrt{4ng_r^2+\Delta^2}-\Delta}{2\sqrt{n}g_r}\right) .
\end{equation}
In the notion $|n,\pm\rangle$, $n=\langle {\sigma}_{2i-1}^{+} {\sigma}_{2i-1}^{-}+{a}_{2i}^\dagger {a}_{2i}\rangle$ refers to the total number of excitations in the unit cell. The corresponding eigenenergies are
\begin{equation}
E_{n,\pm}=(n-\frac{1}{2})\omega_c\pm\frac{1}{2}\sqrt{4ng_r^2+\Delta^2} ,\label{eq:En}
\end{equation}
where $\Delta=\omega_z-\omega_c$ is the detuning between the qubit and the cavity.
At $\Delta=0$, the doublets become
\begin{equation}
|n,\pm\rangle  = \frac{1}{\sqrt{2}} |n,\downarrow\rangle \pm \frac{1}{\sqrt{2}} |n-1,\uparrow\rangle ,
\label{eigenstate2}
\end{equation}
and $E_{n,\pm}=(n-1/2)\omega_{c}\pm\sqrt{n}g_{r}$.

The lowest energy to excite one polariton in the JC model is $\Delta E_{1}=E_{1,-}-E_{0}$,  and the energy to add the second polariton is $\Delta E_{2}=E_{2,-}-E_{1,-}$. At $\Delta=0$, $\Delta E_{1}=\omega_{c}-g_{r}$ and $\Delta E_{2}=\omega_{c}-(\sqrt{2}-1)g_{r}$. In contrast to the excitations in a bare cavity, the JC model has an intrinsic nonlinearity that can be compared to the onsite interaction in the BH model.~\cite{Greentree:2006} For the lower polariton states $\{|n-\rangle\}$ at $\Delta=0$, the effective on-site interaction $U\sim(2-\sqrt{2})g_{r}$, determined by the qubit-cavity coupling. Similar result can be derived at finite detuning.

\subsection{Quantum phases at integer filling}

In the MCJC model with finite $g_{l}$, the polariton excitations can tunnel between adjacent unit cells. The competition between the tunneling and the effective onsite repulsion strongly affects physical properties of this model. The total number of polaritons in the entire lattice
\begin{equation}
N_{t}=\sum_i\left({\sigma}^{+}_{2i-1}{\sigma}_{2i-1}^{-}+{a}^\dag_{2i}{a}_{2i}\right) \label{eq:Nt}
\end{equation}
is a conserving operator. It commutes with the model Hamiltonian, $[N_{t},\,H_{t}]=0$.

We first consider the phase at an integer filling, where $N=\langle N_{t} \rangle $ is an integer multiple of the number of unit cells $L$.
In the limit $g_{l}=0$, the unit cells are decoupled, and the ground state is a product state with all the unit cells being in the lower polariton states $|1,-\rangle$ and the corresponding energy is $E=N E_{1,-}$. It is in the MI phase, which has a finite energy gap for adding or removing one polariton excitation from the system.

To analyze the MCJC model at finite $g_l$, we adopt the polariton mapping technique~\cite{Angelakis:2007,  Koch:2009} and define a polariton operator at each unit cell
\begin{equation}
p^{i}_{n\alpha}\equiv |0,-\rangle_i\langle n,\alpha| .
\end{equation}
This is an operator to annihilate a $n$-polariton state at the $i$th unit cell. When $i\ne j$, $[p^i_{n\alpha}, p^{j\dag}_{m\beta}]=0$. But $p^i_{n\alpha}$ and $p^{i\dag}_{m\beta}$ in the same unit cell do not satisfy the bosonic commutation relation. Using these operators, the spin and photon operators can be expressed as
\begin{align}
{a}_i&=\sum_{n,\alpha,\alpha^\prime}t^n_{\alpha\alpha^\prime} p^{i\dag}_{(n-1)\alpha^\prime}p^i_{n\alpha} ,\\
{\sigma}^-_i&=\sum_{n,\alpha,\alpha^\prime} k^n_{\alpha\alpha^\prime}p^{i\dag}_{(n-1)\alpha^\prime}p^i_{n\alpha} ,
\end{align}
where the coefficients are given by
\begin{eqnarray}
t_{\pm+}^n&=&\sqrt{n}\gamma_{n\pm}\gamma_{(n-1)+}+\sqrt{n-1}\rho_{n\pm}\gamma_{(n-1)-} ,\\
t_{\pm-}^n&=&\sqrt{n}\gamma_{n\pm}\rho_{(n-1)+}+\sqrt{n-1}\rho_{n\pm}\rho_{(n-1)-},
\end{eqnarray}
and $k^n_{\alpha\alpha^\prime} =\rho_{n\alpha}\gamma_{(n-1)\alpha^\prime}$.

The Hamiltonian, $H_{t}$, can be represented in terms of the polariton operators. For example, at $\Delta=0$,
\begin{eqnarray}
H_{t}&=&\sum_{i\alpha n} \left[(n-1/2)\omega_c+\alpha\sqrt{n}g_r\right] p_{n\alpha}^{i\dag}p_{n\alpha}^i \nonumber\\
&&+g_l\sum_{inm}\sum_{\alpha \alpha^{'} \beta \beta^{'}} k_{\alpha\alpha^\prime}^nt^m_{\beta\beta^\prime} H_{hop}^{i} , \label{re_hamiltonian}
\end{eqnarray}
where
\begin{equation}
H_{hop}^{i}=p^{i\dag}_{n\alpha} p^{(i-1)\dag}_{(m-1)\beta^\prime}p^{i-1}_{m\beta}p^{i}_{(n-1)\alpha^\prime}+h.c..
\end{equation}
The first term in Eq.~(\ref{re_hamiltonian}) describes the local polariton states with a nonlinear spectrum that resembles the effective onsite interactions. The second term in Eq.~(\ref{re_hamiltonian}) can be viewed as the hopping of a polariton from site $i-1$ (reducing the number of polaritons from $m$ to $m-1$) to site $i$ (increasing the number of polaritons from $n-1$ to $n$). The competition between these two terms leads to a phase transition between the MI and SF phases. When $g_{l}=0$, the system is in the MI phase. Increasing $g_{l}$, especially in the parameter range where $g_{l}$ becomes comparable to $g_{r}$, the hopping can effectively lower the on-site interaction generated by double occupancy and drive the system into the SF phase at a critical point $g_{l}/g_{r}=\beta_{0}$. This is similar to the phase transition in the BH model with the increase of the hopping integral.~\cite{KuhnerPRB2000} By further increasing $g_{l}$, another MI phase whose properties are similar to the first one by simply exchanging $g_{l}$ with $g_{r}$ emerges above a critical point $g_{l}/g_{r}=1/\beta_{0}$. This analysis agrees with the result obtained by an exact diagonalization calculation~\cite{Seo2015:1}.

\subsection{Quantum phases at half filling}

In the limit there is no coupling between different unit cells, i.e. $g_{l}=0$, at the integer filling with $N=L$, the ground state is in the MI phase with one occupation in the lower polariton state $|1,-\rangle$ per site. We denote this state as $|1 1 1 1 ... \rangle$ and the empty site as $|0\rangle$ for simplicity. In the same limit but away from integer filling, the ground states become highly degenerate. For example, in a lattice of $L=4$ at half filling, i.e., $N=2$, the ground states are six-fold degenerate, with the following polariton configurations $|1010\rangle$, $|0101\rangle$, $|1100\rangle$, $|0110\rangle$, $|0011\rangle$, and $|1001\rangle$. In the weak intercell coupling limit, $g_{l}\ll g_{r}$, we can treat $H_{l}$ as a perturbation. In the first order approximation, the perturbed Hamiltonian can be expressed in this six-fold degenerate subspace as
\begin{equation}
H_{l}=-\frac{g_l}{2}\left(
\begin{array}{cccccc}
0       &0     &1       &1      &1      &1\\
0       &0     &1       &1      &1      &1\\
1       &1     &0       &0      &0      &0\\
1       &1     &0       &0      &0      &0\\
1       &1     &0       &0      &0      &0\\
1       &1     &0       &0      &0      &0
\end{array}\right).\label{pmatrix}
\end{equation}
By diagonalizaiton, we obtain the ground state
\begin{eqnarray}
|g\rangle&=& \frac{\sqrt{2}}{4}\left(|1100\rangle+|0110\rangle+|0011\rangle+|1001\rangle\right) \nonumber \\
&&+\frac{1}{2}\left(|1010\rangle+|0101\rangle\right),\label{perturbation}
\end{eqnarray}
and the corresponding correction to the ground-state energy $E^{(1)}=-\sqrt{2}g_l$.

\section{Method}

We use the finite-lattice algorithm of the DMRG to characterize the critical behavior of the MCJC model.~\cite{White1992} This method has already been used to study critical behavior and quantum dynamics of low-dimensional strongly-correlated systems, including the 1D BH and 1D CCA models.~\cite{KuhnerPRB1998, KuhnerPRB2000, RossiniPRL2007, DSouzaPRA2013} In the calculation, we limit the number of photon excitations at each cavity to be less than or equal to five. For the cases we have examined, we find that this is a good approximation because the contribution from the states with more than five photons at one cavity to the ground-state energy is much smaller than the truncation error.

\begin{figure}
\includegraphics[width=6cm, clip]{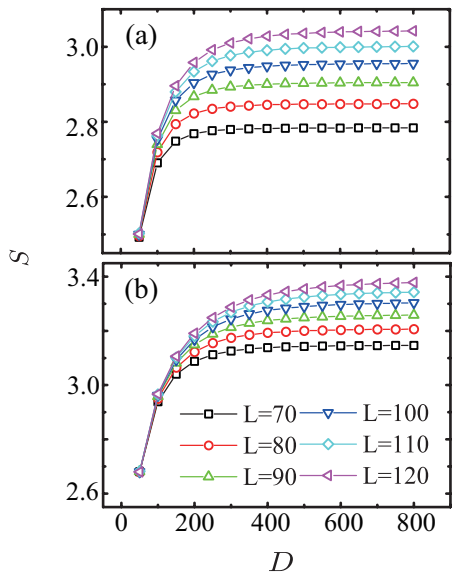}
  \caption{Entanglement entropy $S$ as a function of the bond dimension $D$ for the MCJC model with $g_l=g_r=0.015$, $\omega_c=1$, and $\Delta=0$, at (a) half-filling, $N/L=0.5$, and (b) integer filling, $N/L=1$.
  }
\label{fig_entangle}
\end{figure}

In order to see how fast the DMRG calculations converge with the bond dimension, we calculate the entanglement entropy of the ground state,~\cite{Schollwock2011} defined by
\begin{equation}
S = -\textrm{Tr}\left({\rho}_L \ln \rho_L \right),
\label{eq:entropy}
\end{equation}
where ${\rho}_L$ is the reduced density matrix for the left half of the lattice. Fig.~\ref{fig_entangle} shows the entanglement entropy as a function of the bond dimension $D$ at both half and integer filling at the symmetric point $g_{l}=g_{r}$. After a rapid increase at small $D$, we find that the entanglement entropy becomes almost saturated, which suggests that the ground state wave function is converged, when $D$ is larger than 500. To further ensure the convergence, we take $D=600$ for all the calculations presented below.

For all the cases we have studied, we find that the ground states are translation invariant, as revealed by the distribution functions of the local excitation densities of qubits and cavities, defined by $\bar{n}_i^q  =  \langle{\sigma}_i^{+}{\sigma}_i^-\rangle$ and $\bar{n}_i^r = \langle {a}^\dagger_i {a}_i\rangle$, respectively. Our calculation indicates that the excitation densities are homogenous on the whole lattice with no obvious fluctuations. For example, for the system with $L=120$ at the integer filling and $\Delta=0$, we find that $\bar{n}_i^q = 0.40$ and $\bar{n}_i^r = 0.60$ when $g_l=g_r=0.015$, and $\bar{n}_i^q=0.46$ and $\bar{n}_i^r=0.54$ when $g_l=0.0055$ and $g_r=0.0245$, on all the lattice sites. One interesting observation is that the difference $|\bar{n}_i^q-\bar{n}_i^r|$ between the qubit and cavity excitation densities in the limit of $g_l\ll g_r$ (and similarly, $g_r\ll g_l$) is smaller than that at $g_l = g_r$. This is because in the limit $g_l\ll g_r$, the system behaves like a chain of uncoupled JC models where the excitation is equally splitted between the qubit and the cavity. Whereas at $g_l= g_r$, the polariton excitations can hop along the lattice, enlarging the splitting of local densities. Similar result is found at half filling.

\section{Phase diagram}

\subsection{Phase boundaries}

We calculate the ground-state energy, $E_L(N)$, of the MCJC model with $L$ unit cells and $N$ polariton excitations using the DMRG. The chemical potential for adding or removing a polariton to the ground state is then determined by the formula
\begin{eqnarray}
\mu_p (N, L)&=&E_L(N+1)-E_L(N) , \label{eq:mup}\\
\mu_h (N, L)&=&E_L(N)-E_L(N-1) .\label{eq:muh}
\end{eqnarray}
$\mu_p$ and $\mu_h$ are the chemical potentials for adding and removing a polariton, respectively.~\cite{KuhnerPRB2000}
To obtain the chemical potentials in the thermodynamic limit, we first calculate these quantities at finite lattice systems with $L$ up to 120, and then perform an extrapolation using the formula
\begin{equation}
\mu_{\gamma}(N, L) = \mu_\gamma +b_\gamma /L,  \quad (\gamma = p, h), \label{eq:extropolation}
\end{equation}
where $\mu_{\gamma}$ is the extrapolated chemical potential in the thermodynamic limit.

\begin{figure}
\includegraphics[width=7.5cm, clip]{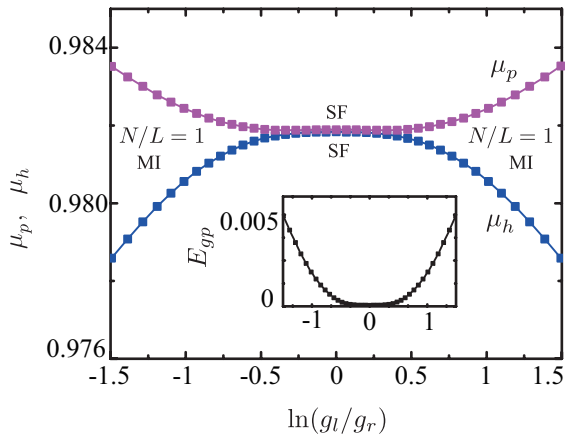}
  \caption{Chemical potentials $\mu_p$ and $\mu_h$ versus $\ln(g_l/g_r)$ for the MCJC model with $g_r+g_l =0.03$, $\omega_c=1$, and $\Delta=0$ at the integer filling $N/L=1$.
  Inset: Energy gap $E_{gp}$ versus $\ln(g_l/g_r)$ for $N/L=1$.  }\label{phase}
\end{figure}

Fig.~\ref{phase} shows the extrapolated chemical potentials, $\mu_p$ and $\mu_h$, as functions of $\ln(g_l/g_r)$. At the integer filling with $N=L$, we find that $\mu_p =\mu_h$ in the regime $g_{l}\sim g_{r}$ (small $|\ln(g_l/g_r)|$) within numerical errors.
As the ratio $|\ln(g_l/g_r)|$ increases, a finite difference appears between $\mu_p$ and $\mu_h$, corresponding to the opening of a finite energy gap for adding or removing a polariton.~\cite{KuhnerPRB2000}
Thus there is a transition from the SF phase in the small $|\ln(g_l/g_r)|$ regime to the MI phase in the large $|\ln(g_l/g_r)|$ regime.
In the SF phase, there is no gap in the energy spectrum and $\mu_p= \mu_h$.
However, in the MI phase, the energy to add or remove a polariton becomes different, and $\mu_p> \mu_h$.
The extrapolated chemical potentials shown in Fig.~\ref{phase} thus determine the phase boundaries that separate the MI and SF phases.
If the chemical potential falls between $\mu_p$ and $\mu_h$ in the MI phase, the filling factor is fixed at $N/L=1$, and the polariton density remains a constant within this phase with zero compressibility.

At the integer filling, there are two quantum critical points which are symmetric with respect to the point $\ln(g_l/g_r) = 0$.
The transitions between the MI and SF phases in this 1D system is of the Kosterlitz-Thouless type~\cite{Kosterlitz_Thouless1973, Kosterlitz1974}.
It results from the competition between $g_l$ and $g_r$.
While in the BH and CCA models, the phase transitions are due to the competition between the hopping and the onsite interaction.~\cite{Fisher:1989, Hartmann:2006, Greentree:2006, Angelakis:2007}

The inset of Fig.~\ref{phase} shows the energy gap for exciting a polariton,  $E_{gp}=\mu_p-\mu_h$, as a function of $\ln(g_l/g_r)$.
It is clear that there is a finite regime in which $E_{gp} =0$. But we cannot accurately determine the positions of the critical points, where $E_{gp}$ begins to deviate from zero, just from the result of this energy gap.

At half filling, the results of chemical potentials indicate that there is no difference in the energy for adding or removing a polariton, i.e. $\mu_p = \mu_h$, in the whole parameter regime.
Thus there is no MI-SF phase transition and the ground state is always in the SF phase.

\subsection{Correlation functions}

The single-particle density matrices for the qubits and cavities are defined respectively by,
\begin{eqnarray}
\Gamma_q(i-j)&=&\langle{\sigma}_{2i-1}^{+}{\sigma}_{2j-1}^-\rangle, \label{gamma2} \\
\Gamma_r(i-j)&=&\langle {a}_{2i}^\dagger{a}_{2j}\rangle. \label{gamma1}
\end{eqnarray}
They measure the correlations of polariton excitations.~\cite{Penrose:1956, Yang:1962}

\begin{figure}
\includegraphics[width=8cm, clip]{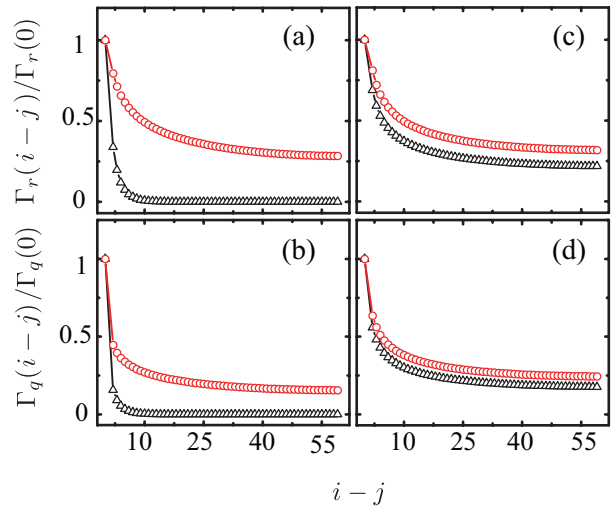}
  \caption{Normalized single-particle density matrices $\Gamma_r(i-j)/\Gamma_r(0)$ and $\Gamma_q(i-j)/\Gamma_q(0)$ versus $i-j$ for the MCJC model with $L=120$ at the integer (a,b) and half (c,d) filling. Circles: $g_l=g_r=0.015$; triangles: $g_l=0.0055, g_r=0.0245$. Other parameters are the same as for Fig.~\ref{fig_entangle}.}
  \label{fig_function}
\end{figure}

Fig.~\ref{fig_function} shows the normalized single-particle density matrices $\Gamma_r(i-j)/\Gamma_r(0)$ and $\Gamma_q(i-j)/\Gamma_q(0)$ versus $i-j$ for two sets of parameters.
At the integer filling, $N/L=1$, both density matrices drop exponentially to zero with the increase of the distance between the two unit cells in the MI phase, $g_l/g_r\ll1$.
In the SF phase, however, the two density matrices remain finite even in the large distance limit, indicating the existence of superfluid off-diagonal long-range order.
At half filling, $N/L=1/2$, the ground state is in the SF phase, and the two density matrices behave qualitatively similar to the case of $g_l=g_r$ at the integer filling: they decrease with the increase of the distance between the two unit cells and saturate to certain finite values in the large distance limit.

\begin{figure}
\centering
\includegraphics[width=6cm, clip]{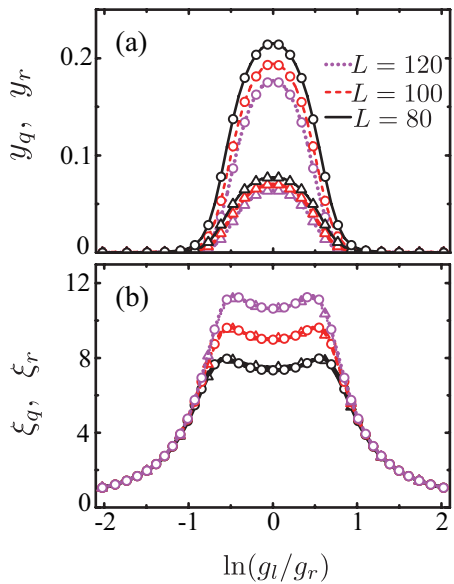}\\
  \caption{(a) Coefficients $y_q$ and $y_r$ and (b) correlation lengthes $\xi_q$ and $\xi_r$ versus $\ln(g_l/g_r)$ for the MCJC model with $L=80, 100, 120$.
  Circles: $y_{r}$ and $\xi_{r}$; triangles: $y_{q}$ and $\xi_{q}$. Other parameters are the same as for Fig.~\ref{fig_entangle}.  }
\label{fig_y}
\end{figure}

The single-particle density matrices can be fitted using the formulae
\begin{equation}
\Gamma_{\alpha}(i-j)= y_{\alpha}+A_q e^{-\frac{|i-j|}{\xi_{\alpha}}}, \quad (\alpha = q, r) , \label{extrapolation}
\end{equation}
where $\xi_q$ ($\xi_r$) is the correlation length between two qubits (cavities).
Fig.~\ref{fig_y}(a) shows the coefficients $y_q$ and $y_r$ as functions of $\ln(g_l/g_r)$.
As expected, both $y_q$ and $y_r$ are zero in the MI phase, but become finite in the SF phase around the regime $\ln (g_l/g_r)$ close to 0.

The correlation lengths, shown in Fig.~\ref{fig_y}(b), increase quickly around the critical points in the MI phases. But we do not see the divergence of the correlation lengths at the critical point, due to the finite lattice size effect. At the critical point, the entanglement entropy is expected to grow logarithmically with the system size.
Thus in order to accurately determine the critical points from the divergent correlation lengths, we need to enlarge not just the lattice size, but also the number of states retained in the DMRG calculation.

\begin{figure}[b]
\includegraphics[width=6.5cm,clip]{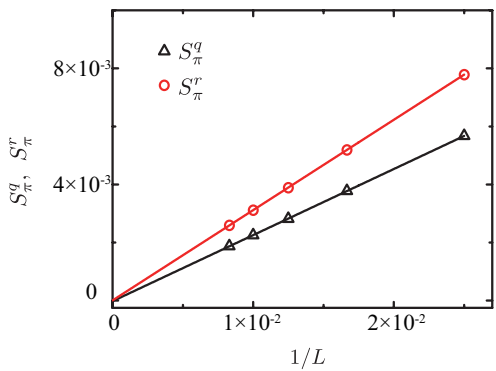}
\caption{Structure factors $S_\pi^q$ and $S_\pi^r$ versus $1/L$ at half filling and $g_l=0.0055, g_r=0.0245$. Other parameters are the same as for Fig.~\ref{fig_entangle}. }
\label{fig_structure}
\end{figure}

From the density-density correlation functions, we calculate the structure factors of the qubits and cavities, defined by
\begin{equation}
S_\pi^{\alpha}=\frac{1}{N^2}\sum_{i,j}(-1)^{|i-j|} \langle {n}_i^{\alpha} {n}_j^{\alpha} \rangle, \quad (\alpha = q, r) , \label{structure}
\end{equation}
where ${n}_i^{r}={a}_{i}^{\dag}{a}_{i}$ and ${n}_i^{q}={\sigma}_{i}^{+}{\sigma}_{i}^{-}$.
Fig.~\ref{fig_structure} shows $S_\pi^{q}$ and $S_\pi^{r}$ as functions of $1/L$ at half filling and $g_l=0.0055$, $g_r=0.0245$. Within numerical errors, the extrapolated structure factors are found to be approximately zero, indicating that there is no crystalline or CDW ordered phases in this system at half filling.~\cite{KuhnerPRB1998}

\subsection{Luttinger parameters}

\begin{figure}[t]
\includegraphics[width=6cm, clip]{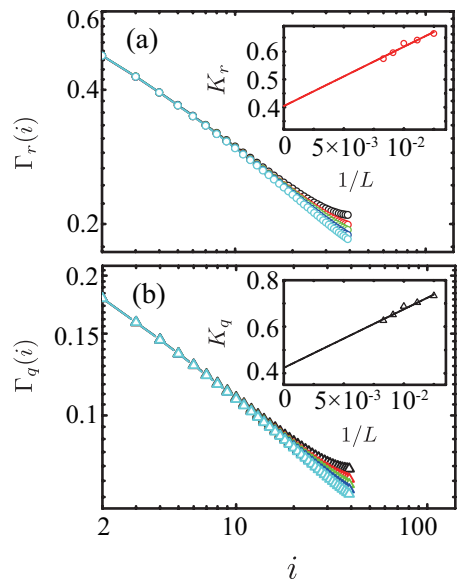}
  \caption{ Log-log plot of the single-particle density matrices $\Gamma_{q, r}(i)$ versus the distance $i$ for the MCJC model at the integer filling $N/L=1$ with $g_l=g_r=0.015$. Insets: $K_{q, r}$ versus $1/L$. The curves are for $L=80,90,100,110,120$ from top to bottom. Other parameters are the same as for Fig.~\ref{fig_entangle}.   }
\label{fig_k}
\end{figure}

In the 1D SF phase, the ground state is a Luttinger liquid,~\cite{Kosterlitz_Thouless1973, Kosterlitz1974} and the correlation functions should decay algebraically with the distance between the unit cells
\begin{equation}
\Gamma_{\alpha}(i-j) \propto |i-j|^{-K_{\alpha}/2}, \quad (\alpha = q, r), \label{power}
\end{equation}
where $K_{\alpha}$ is the Luttinger parameter.
Our calculation, as shown in Fig.~\ref{fig_k}, has indeed confirmed this power law dependence for the single-particle density matrices.
The deviation from the power-law dependence at large distance results from the finite-size effect.
The power-law behaviors have also been observed at half filling, where the ground state is always in the SF phase.
In contrast, in the MI phase, the single-particle density matrices decay exponentially with the distance.

$K_q$ and $K_r$ are important parameters for characterizing the critical behavior of the MCJC model.
To obtain the Luttinger parameters in the thermodynamic limit, an extrapolation of the finite lattice results to the $L\rightarrow \infty$ limit is needed.
From the numerical results, we find that $K_\alpha$ ($\alpha = q, r$) can be well fitted by the formula (see the insets of Fig.~\ref{fig_k})
\begin{equation}
K_{\alpha} = K^0_{\alpha} +\lambda_{\alpha}/L ,\label{Kextra}
\end{equation}
where $K^0_{\alpha}$ is the extrapolated Luttinger parameter in the limit $L \rightarrow \infty$ and $\lambda_{\alpha}$ is a coefficient.

\begin{figure}[t]
\includegraphics[width=7.5cm, clip]{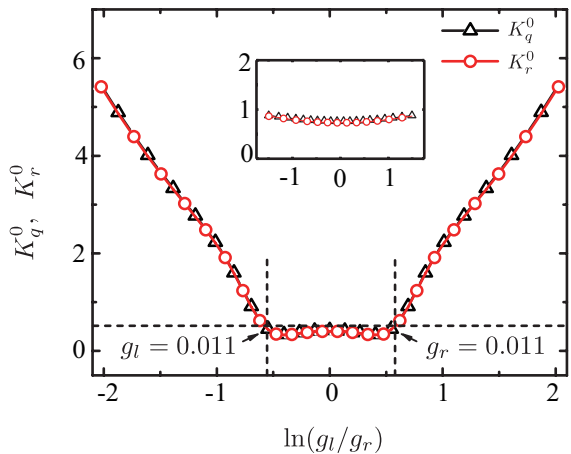}
\caption{$K_q^{0}$ and $K_r^0$ versus $\ln(g_l/g_r)$ at $N/L=1$. The dashed horizontal line is $K_{\alpha}^{0}=1/2$. Inset: $K_{q, r}^{0}$ at $N/L=1/2$. $g_l + g_r =0.03$. Other parameters are the same as for Fig.~\ref{fig_entangle}.}
\label{fig_k1}
\end{figure}

The MI-SF phase transition at integer fillings is of the Kosterlitz-Thouless type, and $K_q^0=K_r^{0}=1/2$ are expected at the critical points for $N/L=1$.~\cite{KuhnerPRB1998, Kosterlitz_Thouless1973, Kosterlitz1974}
In contrast, the corresponding commensurate-uncommensurate phase transitions satisfy $K_q^0 = K_r^0 =1$.
From the calculation of these Luttinger parameters, we can accurately determine the critical points.
Fig.~\ref{fig_k1} shows $K_q^0$ and $K_r^0$ as functions of $\ln(g_l/g_r)$ for the MCJC model at the integer filling $N/L=1$.
Within numerical accuracy, we find that $K_{q}^{0}=K_{r}^{0}$ in the whole parameter range we have studied.
Deep in the SF phase, $g_{l}\sim g_{r}$, $K_{q, r}^{0}<1/2$, indicating that the spatial correlation decreases slowly with the distance.
As $|\ln(g_l/g_r)|$ is increased, $K_{q, r}^{0}$ increase and cross the points $K^0_{q, r}=1/2$ at $\beta_{0}=g_{l}/g_{r}\approx 0.579$ and $1/\beta_{0}=g_{l}/g_{r}\approx 1.727$. The Luttinger parameters at half filling, as shown in the inset of Fig.~\ref{fig_k1}, are found to be $1/2< K_\alpha^0<2$, indicating no phase transition in the whole parameter regime studied.

\section{Conclusions}

To conclude, we study the ground-state properties of the 1D MCJC model of qubits coupled with cavities using the DMRG. The phase boundaries of this model yield symmetric Mott lobes separated by a SF phase due to the intrinsic symmetry between the left and right qubit-cavity interactions. The correlation lengths and the Luttinger parameters at integer and half-integer fillings are determined. By extrapolating the Luttinger parameters to the thermodynamic limit, we accurately determine the critical points separating the MI and SF phases. The structure factors reveal no evidence of crystalline or CDW order at half filling. Our result sheds light on the understanding of the critical behavior of strongly correlated polaritons in the MCJC lattice and related models, such as the CCA, and may stimulate further studies on various stationary and nonequilibrium properties in these systems.

\acknowledgments
J.X. and T.X. are supported by the National Natural Science Foundation of China (Grants No. 11474331 and No. 11190024).  K.S and L.T. are supported by the National Science Foundation under Award No. NSF-DMR-0956064 and UC Multicampus-National Lab Collaborative Research and Training under Award No. LFR-17-477237.

\end{document}